\documentclass[final]{aipproc}
\input{aipcheck}
\layoutstyle{6x9}

\RequirePackage{xspace}
 
\usepackage{relsize}
\def\babar{\mbox{\slshape B\kern-0.1em{\smaller A}\kern-0.1em
    B\kern-0.1em{\smaller A\kern-0.2em R}}}
\newcommand{\Bfactories}{$B$ factories}
\newcommand{\Dzero}{D\O}
\def\Im{\ensuremath{\mathrm{Im}}}
\def\jpsi{\ensuremath{{J\mskip -3mu/\mskip -2mu\psi\mskip 2mu}}\xspace}
    
\begin{document}

% put your own definitions here:
\def\Bz    {\ensuremath{B^0}}
\def\B     {\ensuremath{B}}
\def\Bbar  {\kern 0.18em\overline{\kern -0.18em B}{}}
\def\Bb    {\ensuremath{\Bbar}}
\def\Bzb   {\ensuremath{\Bbar^0}}
\def\Bz    {\ensuremath{B^0}}
\def\Dm    {\ensuremath{{\rm \Delta}m}} 
\def\Dt    {\ensuremath{{\rm \Delta}t}}
\newcommand{\Bzero}{\ensuremath{B^0}}
\newcommand{\Bzbar}{\ensuremath{\bar{B}^0}}
\newcommand{\Btag}{B_{\mathrm{tag}}} 

\newcommand{\epjBase}        {Eur.\ Phys.\ Jour.\xspace}
\newcommand{\jprlBase}       {Phys.\ Rev.\ Lett.\xspace}
\newcommand{\jprBase}        {Phys.\ Rev.\xspace}
\newcommand{\jplBase}        {Phys.\ Lett.\xspace}
\newcommand{\npBase}         {Nucl.\ Phys.\xspace}
\newcommand{\jprl}[1]{\jprlBase\ \textbf{#1}}
\newcommand{\plb}       [1]  {\jplBase\ B~{\bf #1}}
\newcommand{\jprd}      [1]  {\jprBase\ D~{\bf #1}}
\newcommand{\npb}       [1]  {\npBase\ B~{\bf #1}}

%   ...
\newcommand{\ttbs}{\char'134}
\newcommand{\AmS}{{\protect\the\textfont2
  A\kern-.1667em\lower.5ex\hbox{M}\kern-.125emS}}

% add words to TeX's hyphenation exception list
\hyphenation{author another created financial paper re-commend-ed Post-Script}

%\begin{document}

\title{Beauty in the Standard Model and Beyond} 


\author{Gabriella Sciolla\\}{
address={Massachusetts Institute of Technology, 26-443\\ 
77 Massachusetts Avenue \\ 
Cambridge, MA 02139}
}

\begin{abstract}
The study of $CP$ violation in the $B$ system allows us to perform 
quantitative tests of the 
$CP$ symmetry in the Standard Model. %  and look for New Physics.
Many precise measurements of the sides and angles of the Unitarity Triangle
used to test the theory are made possible by 
the abundant experimental data accumulated at the \Bfactories\ and the Tevatron.  
I review the  Standard Model description of $CP$ violation and the key 
measurements which allow us to use $CP$ violation studies as a probe for New Physics.
% look for New Physics in $CP$ violation. 
\end{abstract}

\maketitle

%%%%%%%%%%%%%%%%%%%%%%%%%%%%%%%%%%%%%%%%%%%%
%% MAINMATTER
%%%%%%%%%%%%%%%%%%%%%%%%%%%%%%%%%%%%%%%%%%%%

\section{Introduction}

$CP$ violation  plays  a fundamental  role in the explanation of the matter-dominated universe~\cite{sakarov}. 
% Unfortunately,  little is known about its origin. 
In the Standard Model, $CP$ violation occurs in weak interactions due to the complex phase in the  
quark mixing matrix, the Cabibbo-Kobayashi-Maskawa (CKM) matrix~\cite{KM}. 
This description of $CP$ 
violation, known as the CKM mechanism, provides an elegant and simple explanation of this phenomenon, 
%is very predictive, 
and is in agreement with the experimental measurements 
in the kaon and  $B$ sectors. 
However, the CKM mechanism fails to account for 
the observed baryon-to-photon density ratio in the Universe. 
This suggests that other sources of $CP$ violation must exist besides 
the CKM mechanism, and that $CP$ violation studies may be used as probes for New Physics. 
The key for these studies is to measure $CP$ violation in channels that are theoretically very well understood in the Standard Model, 
and look for deviations from the expectation.

\begin{figure}[b]
  \includegraphics[height=0.16\textheight]{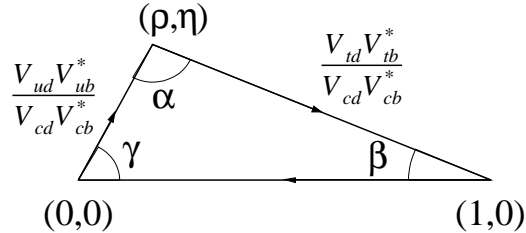}
  \caption{The Unitarity Triangle.}
   \label{UT}
\end{figure}

A convenient tool for these studies is given by the Unitarity Triangle (UT), illustrated in Fig.~\ref{UT}. 
All sides and angles of the UT can be measured in the study of $B$ decays: the time-dependent $CP$ asymmetries 
measure the angles, while the sides can be determined by the measurements
of the semileptonic $B$ decays and the $B$ mixing.  
Since one of the sides of the UT is normalized to a known quantity, only two measurements are necessary to define the triangle 
(e.g. the two sides). Any additional measurement (e.g. an angle) can therefore be used to test the CKM mechanism: 
any inconsistency can be interpreted as a sign of New Physics. Alternatively, we can look for New Physics by measuring 
the same quantity (an angle or a side) through channels that have different sensitivity to New Physics.
%: again, any inconsistency would signal physics beyond the Standard Model. 
% 
It is important to note that  precision and redundancy are essential 
for testing the theoretical predictions. 

The asymmetric \Bfactories\ at SLAC and KEK were specifically designed for such measurements. 
In these machines, electrons and positrons collide at $\sqrt{s}\approx$ 10 GeV and produce an $\Upsilon(4S)$ resonance which decays into a $B\overline{B}$ pair.  The clean environment, typical of $e^+e^-$ colliders, allows  the two 
experiments, \babar\ 
%\cite{babar}  
and Belle, 
%\cite{belle}
to reconstruct $B$ decays with  very high purity and reconstruction efficiency.

The two Tevatron experiments, CDF
%\cite{CDF} 
and \Dzero, %\cite{D0},
 study $B$ hadrons produced in $p\overline{p}$ collisions at $\sqrt{s}\approx$ 2 TeV. 
The harsher experimental environment is compensated by the 
large boost of the $B$ mesons in the laboratory frame
and the fact that all $B$ hadrons can be produced.  
The Tevatron $B$ physics program is therefore complementary to the programs at the \Bfactories.

\section{ MEASUREMENT OF THE ANGLES }

At the \Bfactories, the angles of the Unitarity Triangle 
can be precisely determined through the measurement of the time dependent $CP$ 
asymmetry, $A_{CP}(t)$: 
\begin{equation}
A_{CP}(t) \equiv \frac{N(\Bzb(t)\to f_{CP}) - N(\Bz(t)\to f_{CP})} {N(\Bzb(t)\to f_{CP}) + N(\Bz(t)\to f_{CP})},  
\label{acpt}
\end{equation}
where $N(\Bzb(t)\to f_{CP})$ is the number of \Bzb\ that decay into the $CP$-eigenstate $f_{CP}$ after a time $t$. 
If only one amplitude contributes to the decay,  $ A_{CP}(t) $ can be written as 
%cosine term vanishes, and the sine 
%coefficient simplifies,  $S_f = Im(\lambda)$. 
%$C=0$ and $S=\Im(\lambda_f)$, 
%thus 
\begin{equation}
A_{CP}(t) = - \eta_f \Im(\lambda)\sin(\Dm t) , 
\label{acpt3}
\end{equation}
where $\Delta m$ is the difference in mass between $B$ mass eigenstates and $\eta_f$ is 
the $CP$ eigenvalue of the final state. 
For some decays, $\Im(\lambda)$ is directly and simply related to an angle of the UT. 
For example, in the decay   $B\to\jpsi K^0$, $\Im \lambda =  \sin2\beta$.  

The measurement of $A_{CP}(t)$ utilizes  decays of the $\Upsilon (4S)$ into two neutral $B$ mesons, 
of which one %($B_{CP}$) 
is completely  reconstructed  into a $CP$ eigenstate, 
while the decay products of the other %($\Btag$) 
identify its flavor at decay time. 
The time $t$ between the two $B$ decays is determined by reconstructing the two $B$ decay vertices. 
The $CP$ asymmetry amplitudes are determined from an unbinned maximum likelihood fit 
to the decay time distributions separately for events tagged as \Bz\ and \Bzb .

\begin{figure}[h]
  \includegraphics[height=0.2\textheight]{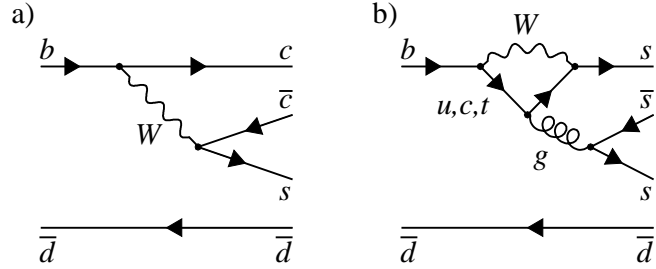}
 \caption{Feynman diagrams that mediate the \Bz\ decays used to measure the angle $\beta$: 
 a) $\Bz\to\mbox{charmonium}+K^0$; b) penguin dominated $B$ decays.}  
 \label{fey}
\end{figure}

%\subsection{$\Bz\to\mbox{charmonium}+K^0$}	
\subsection{Measurement of the angle $\beta$}	
%           -----------------------------
The decays $\Bz\to\mbox{charmonium}+K^0$, 
known as ``golden modes'' for the measurement of the angle $\beta$, 
are dominated by a tree level diagram $b\to c\overline{c}s$ with 
internal $W$ boson emission (Fig.~\ref{fey}a). 
Besides the theoretical simplicity, 
these modes 
%also offer experimental advantages 
are advantageous because of  their  
relatively large branching fractions ($\sim 10^{-4}$) 
and the presence of the narrow $\jpsi$ resonance in the final state,
which provides a powerful rejection of combinatorial background.
The $CP$ eigenstates considered for this analysis are $\jpsi K_S$, $\psi$(2S)$K_S$, 
$\chi_{c1}K_S$, $\eta_cK_S$ and $\jpsi K_L$. 

\begin{figure}[th]
 \includegraphics[height=0.35\textheight]{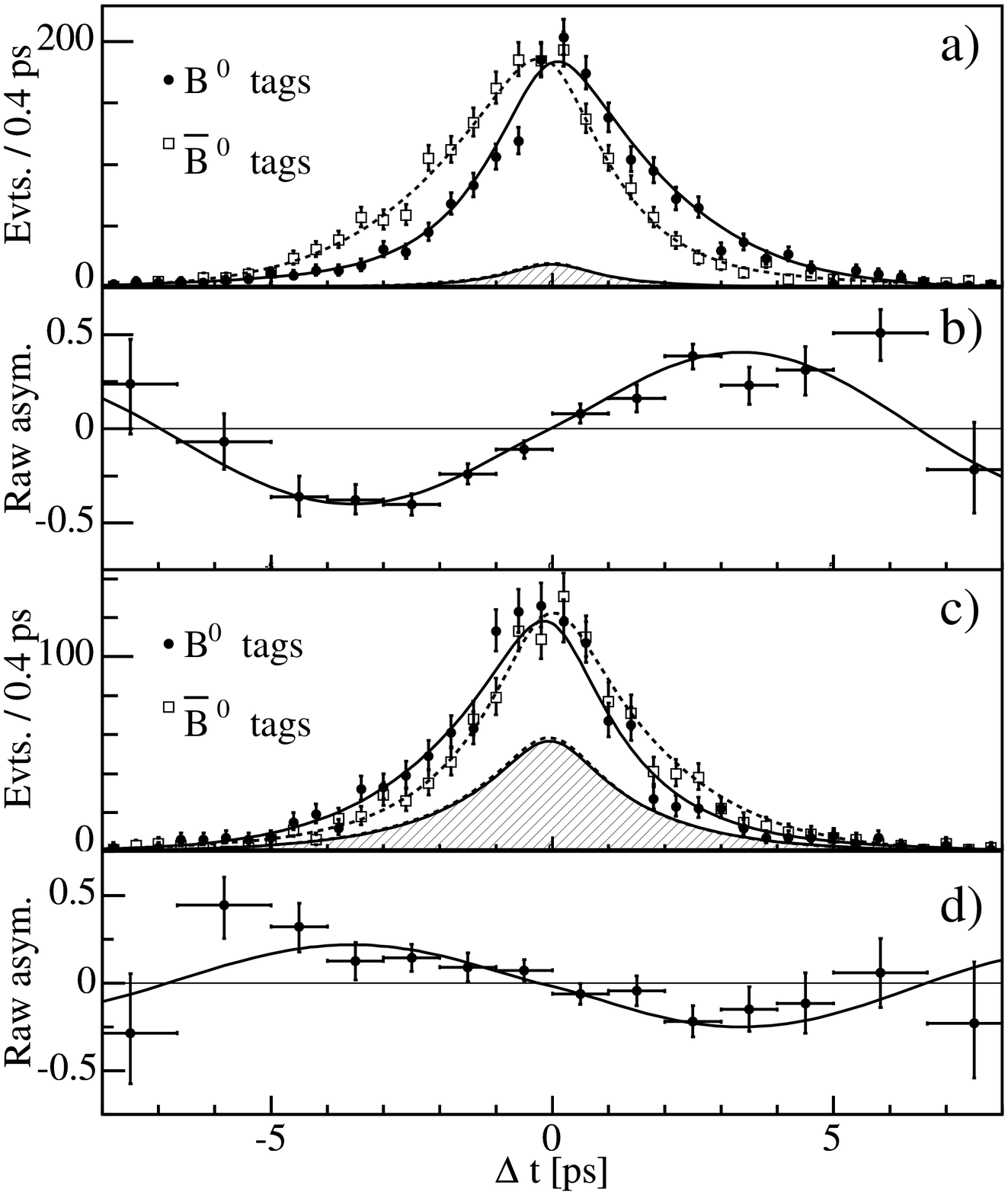}
 \includegraphics[height=0.4\textheight]{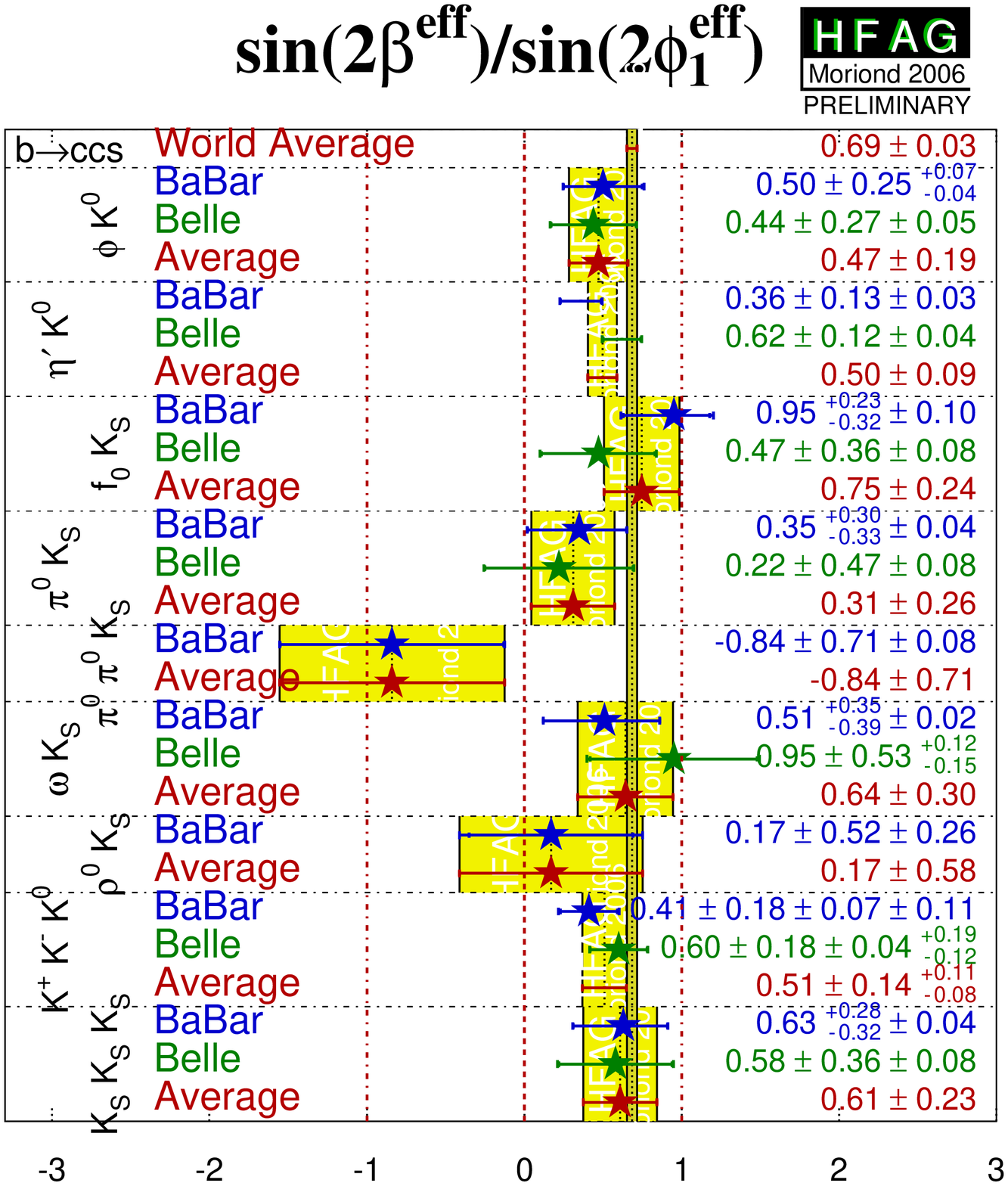}
 \caption{Left: \babar\ measurement of $\sin 2\beta$ in the ``golden modes''. 
Plot a) shows the time distributions for events tagged as \Bz\ (full dots) or \Bzb\ (open squares) 
in $CP$ odd  (charmonium $K_S$) final states. Plot b) shows the  
corresponding raw $CP$ asymmetry with the projection of the unbinned maximum likelihood fit superimposed. 
Plots c) and d) show  the corresponding distributions for $CP$ even ($\jpsi K_L$) final states. 
Right: measurements of $\sin2\beta$ in penguin dominated modes. 
}
%\vspace{-0.4cm}
\label{sin2b}
\end{figure}

The results for the measurements of $CP$ violation 
in $\Bz\to\mbox{charmonium}+K^0$ are illustrated in Fig.~\ref{sin2b} (left). 
The asymmetry between the \Dt\ distributions of events tagged 
as $B^0$ and events tagged as $\Bzb$,  
clearly visible in a) and c),  is  
a striking manifestation of $CP$ violation in the $B$ system. 
The corresponding time dependent  $CP$ asymmetry 
is shown in b) and d). 
\babar\ measures 
$\sin 2\beta=0.722\pm 0.040\pm 0.023$~\cite{sin2bbabar}. 
When combining this result with the corresponding measurement from the Belle experiment  
$\sin 2\beta=0.652\pm 0.039\pm 0.020$~\cite{sin2bbelle}, we obtain  
$\sin 2\beta=0.685\pm 0.032$~\cite{hfag}. 
This implies that the angle $\beta$ is known to a precision of 1 degree. 

% % 

% \subsection{$\beta$ from penguin dominated $\Bz$ decays }
%           -----------------------------
In the Standard Model, 
final states dominated by $b\to s \overline{s} s $ or $b\to s \overline{d} d $ decays 
offer a clean and independent way of measuring $\sin2\beta$~\cite{sPenguin}. 
Examples of these final states are 
$\phi K^0$,  $\eta 'K^0$, $f_0K^0$, $\pi^0 K^0$, $\omega K^0$, $K^+K^-K_S$ and  $K_S K_S K_S$.
These decays are mediated by the gluonic penguin diagram illustrated in Fig.~\ref{fey}b. 
With contributions from 
physics beyond the Standard Model, new particles such as  
squarks and gluinos could participate in the loop and affect the time 
dependent asymmetries~\cite{phases}.  

The decay $\Bz\to \phi K_S$ is ideal for these studies. In the Standard Model, 
this decay is an almost pure $b\to s \overline{s} s $ penguin decay, and its $CP$ asymmetry is 
expected to coincide with the one measured in charmonium + $K^0$ decays within a few percent~\cite{phases}.
Experimentally, this channel is also very clean,  
thanks to the powerful background suppression due to the narrow $\phi$ resonance. Unfortunately, 
the branching fraction for this mode is quite small ($\approx 8\times 10^{-6}$), therefore the 
measurement is limited by a large statistical error. 

The decay $\Bz\to \eta^{\prime}K_S$ is favored by a
larger branching fraction ($\approx 6\times 10^{-5}$).  
In the Standard Model, this decay is also dominated by penguin diagrams; other contributions 
are expected to be small~\cite{etaprimetheory}.  

A summary of the measurements of $A_{CP}(t)$ in penguin modes~\cite{phiks,etaprimeKs,kspi0,f0k0,omegaKs,sin2bbelle}
by the \babar\ and Belle experiments is reported in Fig.~\ref{sin2b} (right). 
The naive averaging of  all the penguin modes~\cite{hfag}
results in a 2.5$\sigma$ deviation  from the value of  $\sin 2\beta$ measured in the golden mode. 
However, this discrepancy  has to be interpreted with caution since each 
mode can be affected by new physics in different ways.

\subsection{ Measurement of the angles $\alpha$ and $\gamma$}

The most accurate determination  of the angle $\alpha$ comes from the 
measurement of the time-dependent $CP$ asymmetry in $\Bz\to\rho^+\rho^-$ decays. 
In the SM, these decays are dominated by a $b\to u \overline{u} d$ tree diagram. 
In the assumption that no other diagram contributes to the final state, $\Im\lambda = \sin2\alpha$. 
Penguin diagrams can contribute to this final state, but their contribution is  thought to be small 
because of the small branching fraction 
measured for the $\Bz\to\rho^0\rho^0$ decay~\cite{rho0rho0}. 
Since $\rho$ is a vector meson, the $\rho^+\rho^-$ final state is characterized by three possible angular momentum states, 
and therefore it is expected to be an admixture of $CP=+1$ and $CP=-1$ states. 
However, polarization studies~\cite{rhorho_babar,rhorho_belle} indicate that this final state is almost completely longitudinally polarized, 
and therefore almost a pure $CP=+1$ eigenstate.  
The parameter $\sin2\alpha$  is therefore measured from the amplitude of the time dependent $CP$ asymmetry, 
using the same technique  described for the measurement of the angle $\beta$. 

Other final states, such as $\Bz\to\pi^+\pi^-$ and $B\to\rho\pi$~\cite{alphababar,alphabelle}, provide additional constraints on the angle $\alpha$.  
Combining  all \babar\  and Belle results, we measure $\alpha =(105^{+15}_{-9})^{\circ}$~\cite{CKMFitter}.

% \subsection{ Measurement of $\gamma$}
The angle $\gamma$ is measured exploiting the interference between the decays $B^+\to D^0K^+$ and $B^+\to\overline{D}{}^0K^+$, where both 
$D^0$ and $\overline{D}{}^0$ decay to the same final state.
This measurement can be performed in three different ways: 
utilizing decays of $D$ mesons to $CP$ eigenstates~\cite{GWL};
utilizing doubly Cabibbo-suppressed decays of the $\overline{D}$ meson~\cite{ADS}; 
exploiting the interference pattern in the Dalitz plot of $D\to K_S\pi^+\pi^-$ decays~\cite{GGSZ}. 
Currently, the last analysis provides the best measurement of the angle $\gamma$. 
Combining all results from \babar\ and Belle, we measure $\gamma=(65\pm 20)^{\circ}$~\cite{utfit}.

\section{ MEASUREMENT OF THE SIDES }

The left side of the Unitarity Triangle is determined by the ratio of the CKM elements $|V_{ub}|$ and $|V_{cb}|$. 
Both elements are measured in the study of semi-leptonic $B$ decays. 
The measurement of $|V_{cb}|$ is already very  precise, with errors of the order of 2\%~\cite{hfag}.  
The determination of $|V_{ub}|$ is more challenging, mainly due to the large background 
coming from $b\to c\ell\nu$ decays, about 50 times more likely to occur than $b\to u\ell\nu$ transitions. 

Two approaches, inclusive and exclusive, can be used to determine $|V_{ub}|$.
In inclusive analyses of $B\to X_u\ell\nu$,  the $b\to c\ell\nu$ background is suppressed by cutting on a number of kinematical variables. 
This implies that only partial rates can be directly measured, and theoretical assumptions are used to infer the total rate and extract $|V_{ub}|$. 
The theoretical error associated with these measurements is $\approx 4\%$. 
Averaging all inclusive measurements from the \babar, Belle, and CLEO experiments we determine 
$|V_{ub}|=(4.45\pm 0.33 ) \times 10^{-3}$~\cite{hfag}, where the error includes statistical, systematic and theoretical errors. 

In exclusive analyses, $|V_{ub}| $ is extracted from the measurement of the branching fraction $B\to \pi\ell\nu$. These analyses are 
usually characterized by a good signal/background ratio, but lead to measurements with large statistical errors due to the the small 
 branching fractions of the mode studied. In addition, the theoretical errors, dominated by the uncertainties in the form factor calculation, 
 are $\approx 12\%$. Both experimental and theoretical errors are expected to decrease in the future, making this approach  competitive 
 with the inclusive method. 
 
The right  side of the Unitarity Triangle is determined by the ratio of the CKM elements $|V_{td}|$ and $|V_{ts}|$. 
This  ratio can be determined with small theoretical uncertainly from the measurement of ratio of the $B^0$ and $B_s$ mixing frequencies. 
While the $B^0$ mixing parameter $\Delta m_d$ has been measured very precisely by 
many experiments~\cite{hfag}, the  $B_s$ mixing parameter $\Delta m_s$ had escaped detection until recently, due to the 
difficulty in detecting its very fast oscillations. 
This spring, the Tevatron experiments succeeded in this endeavor,  and  published evidence for 
$B_s$ oscillations~\cite{D0mixing,CDFmixing}, as described in detail in~\cite{cdftalk}.  
%
%At the Tevatron, the $B_s$ mesons are exclusively reconstructed in their hadronic   or semileptonic  decays. Their 
% flavor at production time is inferred by tagging the flavor of the other B hadron produced in the opposite hemisphere, or by looking 
 %at the sign of fragmentation kaons produced in the same hemisphere.  The time between production and decay of the  $B_s$ mesons is 
% then  determined from the measurements of the boost of the  $B_s$  meson and the distance between  
% the interaction point and the B meson decay vertex. 
%
The value of  $\Delta m_s $ measured by  CDF is $17.33^{+0.42}_{-0.21}\pm 0.07 \mathrm{ps}^{-1}$. Combining this measurement
 with    the world average for  $\Delta m_d$,  
one can extract $|V_{td}/V_{ts}|=0.208^{+0.008}_{-0.007}$.

\section{ CONCLUSION }

\begin{figure}[h]
 \includegraphics[height=0.5\textheight]{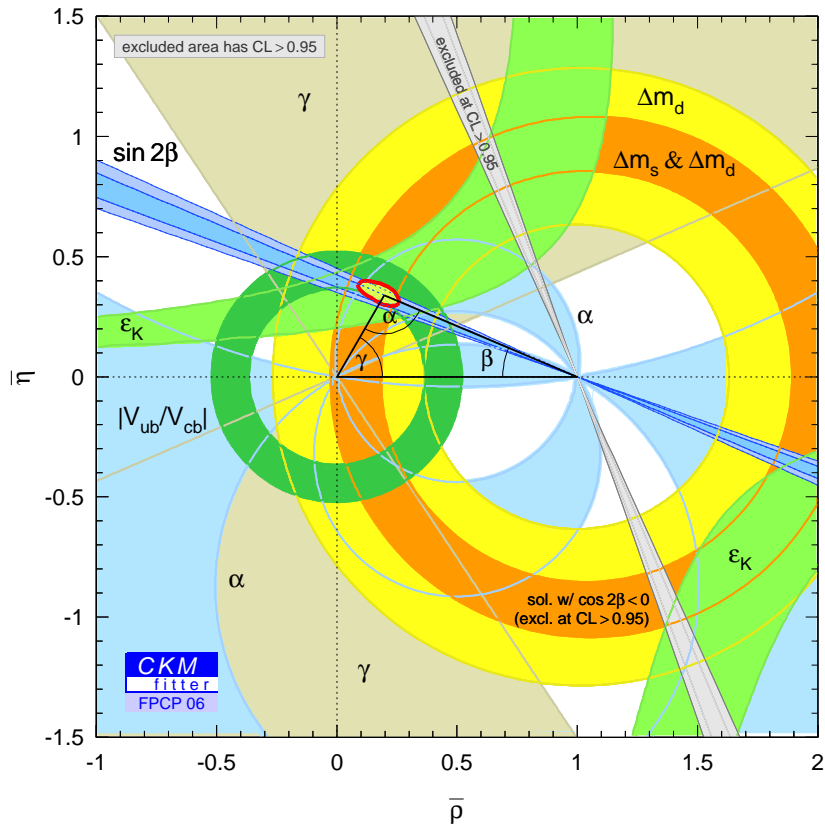}
 \caption{  Constraints on the apex of the Unitarity Triangle resulting from the various measurements 
                   of its  sides and angles. }  
 \label{rhoeta}
\end{figure}

Precise and redundant measurements of the sides and angles of the Unitarity Triangles 
have provided a crucial test of  $CP$ violation in the Standard Model. 
The constraints on the ($\rho ,\eta$) plane  due to the 
measurements described in this article are illustrated in Fig.~\ref{rhoeta}. 
The comparison shows excellent agreement between all  
measurements, as predicted by the  CKM mechanism.

Measurements of time-dependent $CP$ violation asymmetries  
in penguin-dominated modes are sensitive to contributions from physics beyond the 
Standard Model. These measurements are still heavily dominated by statistical errors and will 
benefit greatly from a factor two increase in statistics that both \babar\ and Belle are planning to achieve by 2008. 

%

%%%%%%%%%%%%%%%%%%%%%%%%%%%%%%%%%%%%%%%%%%%%%%%%
%% BACKMATTER
%%%%%%%%%%%%%%%%%%%%%%%%%%%%%%%%%%%%%%%%%%%%%%%%

\bibliographystyle{aipproc}   % if natbib is available

\bibliography{sample}

\IfFileExists{\jobname.bbl}{}
 {\typeout{}
  \typeout{******************************************}
  \typeout{** Please run "bibtex \jobname" to optain}
  \typeout{** the bibliography and then re-run LaTeX}
  \typeout{** twice to fix the references!}
  \typeout{******************************************}
  \typeout{}
 }

\end{document}